\documentclass[10pt,twocolumn,twoside,english]{IEEEtran}
\usepackage{color}

\usepackage{cite}
\ifCLASSINFOpdf
   \usepackage[pdftex]{graphicx}
\else
  \usepackage[dvips]{graphicx}
   \DeclareGraphicsExtensions{.eps}
\fi
\usepackage{subfigure}

\usepackage{amsmath}
\usepackage{amsfonts,helvet}
\usepackage{amssymb}

\usepackage{algorithm}
\usepackage{algpseudocode}
\usepackage{array}
\usepackage{multirow}

\usepackage{hyperref}

\begin{document}
\title{Nonlinear Self-Interference Cancellation for Full-Duplex Radios: From Link- and System-Level Performance Perspectives}

\author{Min Soo Sim,~\IEEEmembership{Student~Member,~IEEE}, 
MinKeun Chung,~\IEEEmembership{Student~Member,~IEEE}, 
Dongkyu Kim,~\IEEEmembership{Member,~IEEE},
Jaehoon Chung, 
Dong Ku Kim,~\IEEEmembership{Senior~Member,~IEEE}, 
and Chan-Byoung~Chae,~\IEEEmembership{Senior~Member,~IEEE}\\
%\thanks{This work was supported by Samsung Electronics.}
\thanks{This article was presented in part at the IEEE Globecom FDWC Workshop 2016~\cite{Sim2016SIC}.

M. S. Sim, M. Chung, D. K. Kim, and C.-B. Chae are with Yonsei University, Korea (E-mail: \{simms, minkeun.chung, dkkim, cbchae\}@yonsei.ac.kr). D. Kim and J. Chung are with LG Electronics (E-mail: \{dkyu.kim, jaehoon.chung\}@lge.com).

This work was in part supported by the the MISP under the ``ICT Consilience Creative Program" (IITP-2015-R0346-15-1008), the ICT R\&D Program of MSIP/IITP (B0126-15-1017), and LG Electronics. 
}}

% The paper headers
%\markboth{Transactions on Wireless Communications,~Vol.~, No.~, Month~Year}%
%{Shell \MakeLowercase{\textit{et al.}}: Massive MIMO Operation in Partially Centralized Cloud Radio Access Networks}

% The only time the second header will appear is for the odd numbered pages
% after the title page when using the twoside option.
% 

% If you want to put a publisher's ID mark on the page you can do it like
% this:
%\IEEEpubid{0000--0000/00\$00.00~\copyright~2012 IEEE}
% Remember, if you use this you must call \IEEEpubidadjcol in the second
% column for its text to clear the IEEEpubid mark.
% make the title area
\maketitle
%%%%%%%%%%%%%%%%%%%%%%%%%%%%%%%%%%%%%%%%%%%%%%%%%%%%%
%%%%%%%%%%%%%%%%%%%%%%%%%%%%%%%%%%%%%%%%%%%%%%%%%%%%%
\begin{abstract}

One of the promising technologies for LTE Evolution is full-duplex radio, an innovation is expected to double the spectral efficiency. To realize full-duplex in practice, the main challenge is overcoming self-interference, and to do so, researchers have developed self-interference cancellation techniques. Since most wireless transceivers use power amplifiers, especially in cellular systems, researchers have revealed the importance of nonlinear self-interference cancellation. In this article, we first explore several nonlinear digital self-interference cancellation techniques. We then propose a low complexity pre-calibration-based nonlinear digital self-interference cancellation technique. Next we discuss issues about reference signal allocation and the overhead of each technique. For performance evaluations, we carry out extensive measurements through a real-time prototype and link-/system-level simulations. For link-level analysis, we measure the amount of canceled self-interference for each technique. We also evaluate system-level performances through 3D ray-tracing-based simulations. Numerical results confirm the significant performance improvement over a half-duplex system even in interference-limited indoor environments.

%In this article, we present the design, implementation, and evaluation of a candidate for next generation cellular systems - full-duplex LTE radios.
\end{abstract}


% For peer review papers, you can put extra information on the cover
% page as needed:
% \ifCLASSOPTIONpeerreview
% \begin{center} \bfseries EDICS Category: 3-BBND \end{center}
% \fi
%
% For peerreview papers, this IEEEtran command inserts a page break and
% creates the second title. It will be ignored for other modes.
\IEEEpeerreviewmaketitle

%\newpage
%%%%%%%%%%%%%%%%%%%%%%%%%%%%%%%%%%%%%%%%%%%%%%%%%%%

%%%%%%%%%%%%%%%%%%%%%%%%%%%%%%%%%%%%%%%%%%%%%%%%%%%
\section{Introduction}

As a solution to the tremendous expansion of mobile traffic, researchers have been developing, over the past several years, fifth generation (5G) wireless communication/Long Term Evolution (LTE) Evolution. One of the main requirements for this service is to provide a 1000-fold improvement in throughput over current fourth generation mobile networks such as LTE Advanced~\cite{5Gforum2016vision}. To achieve this requirement, researchers have strived to accomplish the following: improving spectral efficiency (bps/Hz), expanding system bandwidth (Hz), and/or increasing throughput per area (bps/$\text{m}^2$). Several promising technologies have been developed to improve spectral efficiency such as massive multiple-input and multiple-output (MIMO), 3-dimensional (3D) beamforming, and in-band full-duplex radios.

In-band full-duplex radios (full-duplex radios hereafter) simultaneously transmit and receive on the same frequency band~\cite{Sim2016SIC, Chung2015Prototyping, Sachin2011Practical, Sachin2013Full, Heino2015FDRelay, Duarte2014, aryafar2012MIDU, Ahmed2015All, Khandani1, Hua1}. Full-duplex systems are expected, by definition, to double the spectral efficiency of half-duplex systems. Current commercial systems, however, have engaged with little attention to this type of system due to its propensity for self-interference. Self-interference is the phenomenon of a signal, transmitted from a transmitter, being received by its own receiver while that receiver is trying to receive a signal sent from another device (signal-of-interest). The self-interference, which is generally far stronger than signal-of-interest, makes it impossible for a device to decode the signal-of-interest. Today, mobile networks, so as to avoid self-interference, operate in half-duplex. Frequency-division duplex (FDD) systems prevent self-interference by allocating different frequency bands for uplink and downlink. Time-division duplex (TDD) systems transmit and receive at different times. To deal with the self-interference issue in full-duplex systems, researchers have developed several self-interference cancellation techniques, the objective of which is to mitigate or cancel the self-interference to noise level.

\begin{figure}[t]
   \centerline{\resizebox{1.0\columnwidth}{!}{\includegraphics{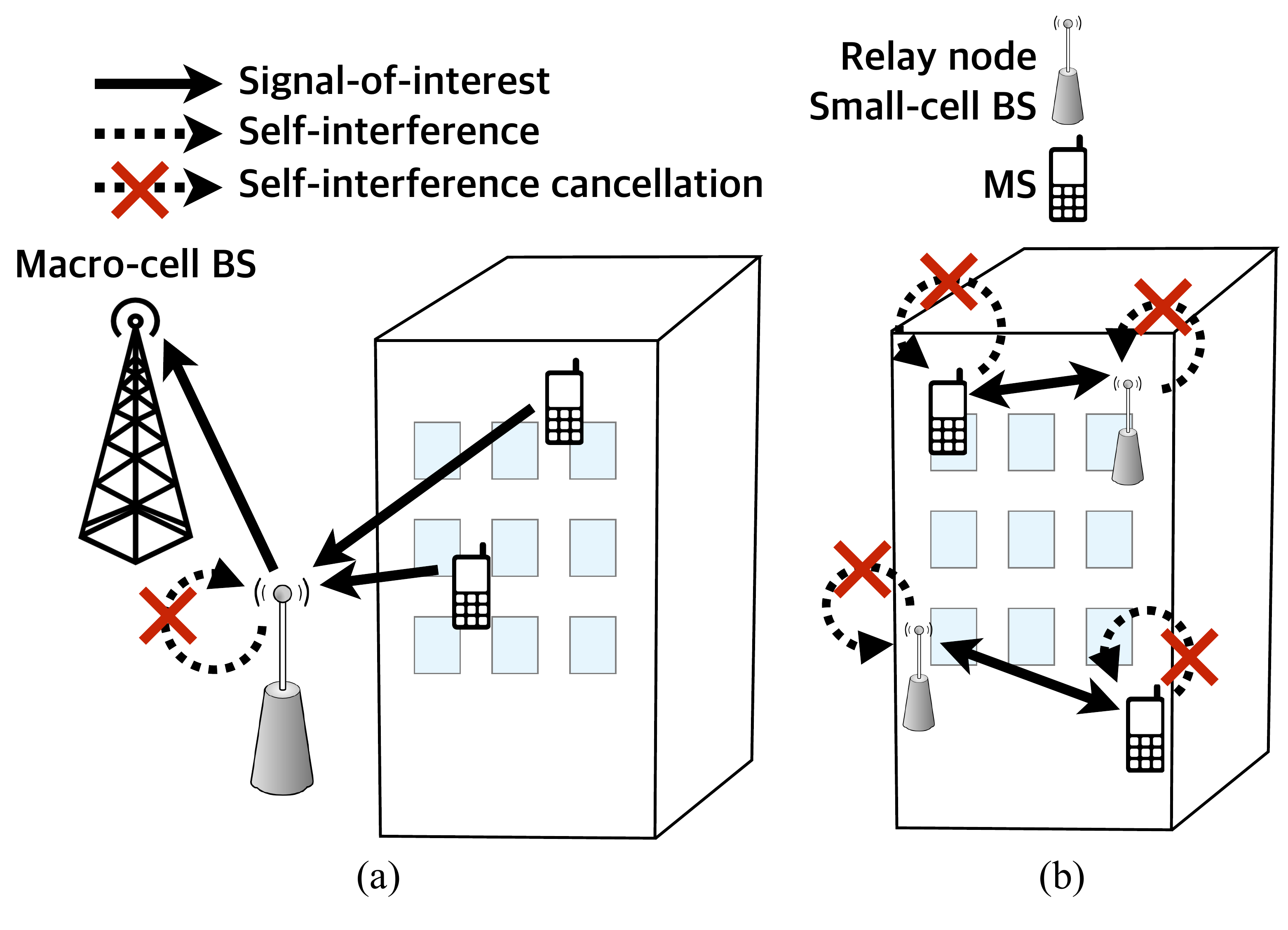}}}
   \caption{Scenarios of full-duplex-radio-based LTE systems. (a) A full-duplex LTE relay system and (b) a full-duplex LTE small-cell system.}
   \label{fig.scenario}
\end{figure}

\begin{figure*}[t!]
   \centerline{\resizebox{1.4\columnwidth}{!}{\includegraphics{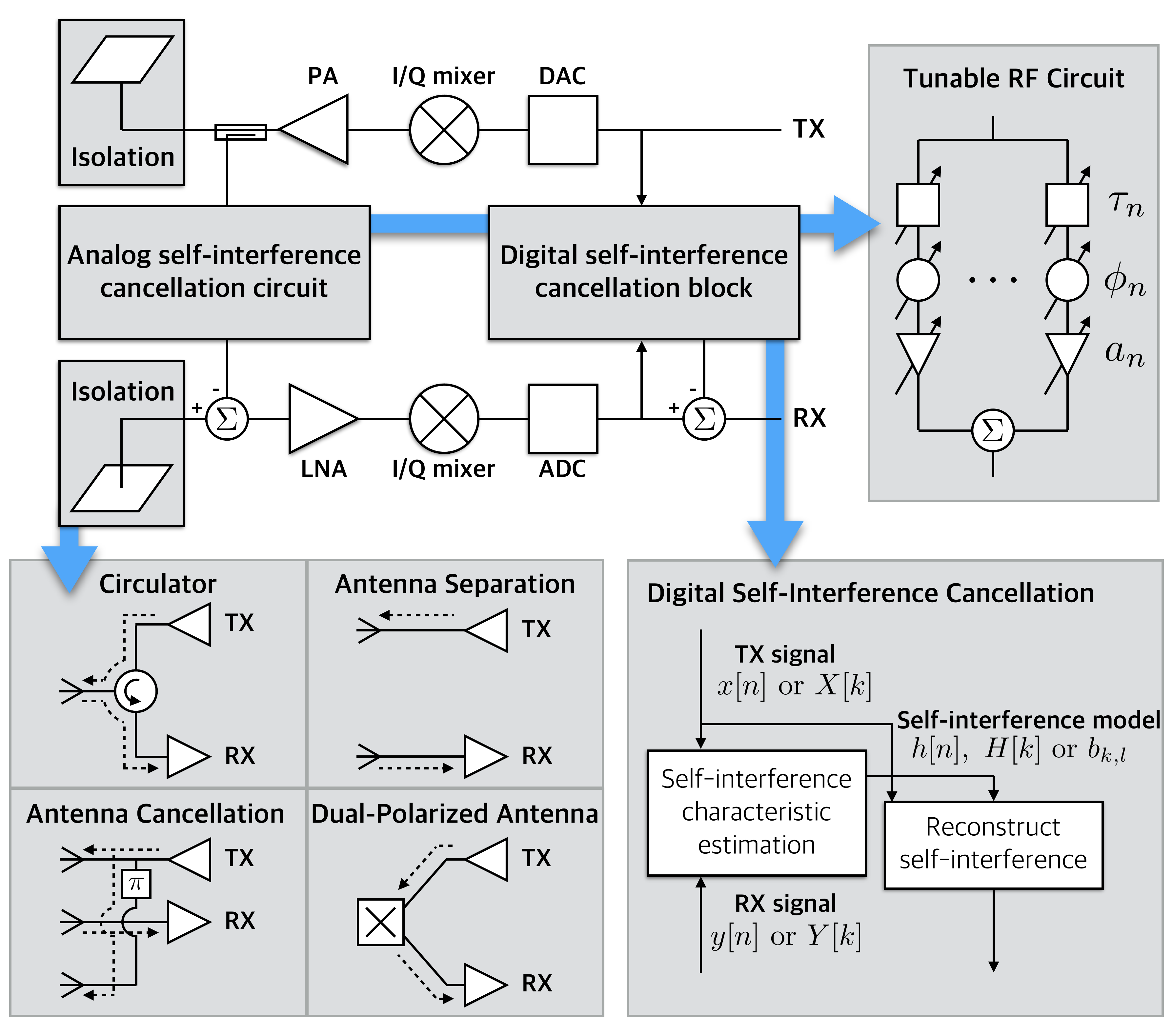}}}
   \caption{A full-duplex transceiver with self-interference cancellation techniques. Representative isolation, active analog self-interference cancellation, and digital self-interference cancellation schemes are illustrated.}
   \label{fig.sic}
\end{figure*}

Applying full-duplex technology, due to its high transmit power, is a challenge in cellular networks such as LTE. Typical full-duplex systems provide self-interference cancellation of approximately 120~dB~\cite{Sachin2013Full, Heino2015FDRelay}. For a macro-cell base station (BS) whose transmission power is up to 46~dBm and a noise level of -90~dBm, it is not feasible to suppress self-interference to the noise level. Therefore, applications of full-duplex radio in LTE systems could be limited to relay systems or small-cell systems where transmission power is at most 23~dBm (see Fig.~\ref{fig.scenario}). 
To apply full duplex to these scenarios, the nonlinearity of power amplifiers becomes a bottleneck of self-interference cancellation.
Power amplifier's nonlinearity is one of the imperfections of radio-frequency (RF) transceivers; other imperfections include in-phase and quadrature (I/Q) imbalances and phase noise. Since these RF imperfections are not sufficiently suppressed in the analog domain, digital self-interference cancellation should share the burden. Most of the digital processing in conventional communication systems, however, is designed linearly. Therefore, special tech that can handle this nonlinearity are required for the processing to perform self-interference cancellation well. 

In this article, we investigate several techniques that can cancel nonlinear self-interference. We introduce the concepts, compare reference signal allocations, and evaluate link- and system-level performances. To the best of our knowledge, this is the first attempt to investigate \emph{system-level performances of full-duplex radios based on measured data}.

The rest of this article is organized as follows. In Section~\ref{overview_SIC}, we introduce several self-interference cancellation techniques including isolation, active analog cancellation, and linear digital cancellation. Section~\ref{nonlinear_SIC} details representative nonlinear digital self-interference cancellation techniques. We investigate link- and system-level performance evaluations in Section~\ref{evaluation}, and our conclusions are given in Section~\ref{conclusion}.

%%%%%%%%%%%%%%%%%%%%%%%%%%%%%%%%%%%%%%%%%%%%%%%%%%%%%%%%%%%%%%%%%%%%%%%%%%

\section{An Overview of Self-Interference Cancellation Techniques}
\label{overview_SIC}

Most prior work on full-duplex radios has focused on designing a wireless transceiver that can perform a sufficient amount of self-interference cancellation. Included in such work are the following: 1) designing an antenna and configuration that lessen coupling between a transmitter and a receiver, 2) suppressing self-interference by mimicking the analog self-interference signal and subtracting from it, and 3) canceling digitalized self-interference by modeling self-interference and subtracting from it. Figure~\ref{fig.sic} illustrates a schematic of several self-interference cancellation technologies. The ultimate goal is to maximize a total cancellation amount by integrating and optimizing these techniques. In this section, to show how self-interference cancellation works in full-duplex systems, we briefly introduce some of the conventional self-interference cancellation techniques.

\subsection{Analog Self-Interference Cancellation}

Analog self-interference cancellation suppresses self-interference in the analog domain, i.e., before the signal passes through an analog-to-digital converter (ADC). The main role of analog self-interference cancellation is to take up a portion of the total cancellation amount and to make sure that the residual self-interference can be canceled out in the digital domain. Since the cancellation performance in the digital domain is limited by the dynamic range of an ADC, analog self-interference cancellation should suppress certain portions of self-interference. According to tunability or adaptability, analog self-interference can be categorized, in this article, as isolation (a.k.a. passive analog self-interference cancellation) and active analog self-interference cancellation.

\smallskip
\subsubsection{Isolation}

Isolation suppresses self-interference signals in the analog domain without any adaptive tuning. Special antenna structures and configurations, and passive devices are employed to weaken, passively, self-interference signals. A basic method for isolation is antenna separation, which causes path-loss between a transmit antenna and a receive antenna~\cite{Duarte2014}. All transceivers that use different antennas for transmitters and receivers can obtain self-interference suppression gain. A circulator was employed for the system that transmitters and receivers share antennas~\cite{Sachin2013Full}. A circulator is a three-port device that transfers a transmit signal from a transmit port to an antenna and a receive signal from an antenna to a receive port while the signal from the transmit port is blocked from the receiver port. With a dual-polarized antenna, a leakage from a transmitter to a receiver can be prevented despite the short physical distance~\cite{Chung2015Prototyping}. 
To obtain extra isolation, a technique called antenna cancellation was proposed, which uses a symmetric antenna configuration and a $\pi$-phase shifter to take advantage of destructive interference~\cite{aryafar2012MIDU}.

\smallskip
\subsubsection{Active Analog Self-Interference Cancellation}

Active analog self-interference cancellation, through adaptive tuning and algorithms, aims at the dominant components of self-interference. The dominant components here represent a line-of-sight component of a system with separate transmit and receive antennas, or a leakage from a circulator of a system with a single transceiver antenna. A typical solution of active analog self-interference is a tunable RF circuit~\cite{Sachin2011Practical, Sachin2013Full}. The circuit, which consists of several taps with RF components such as attenuators, phase shifters, and delays, uses a replica of the transmitted signal as an input, and tries to mimic the self-interference signal. Since the circuit directly exploits the transmitted signal's RF imperfections, which in the digital domain is in fact hard to handle, it would have a better chance of suppressing the self-interference in analog cancellation. To follow the time-varying characteristic of the self-interference channel, a real-time adaptive algorithm and extra reference signal are needed to control RF components. 
% The number of taps and parameters of RF components are properly determined depending on the hardware characteristic of the system and the target self-interference components to suppress. Note that having more taps or controllable RF components implies accurate self-interference suppression, but also increases computational complexity. 
Note that minimizing a noise of the RF circuit, which can be achieved by employing passive devices that generate virtually no noise, is essential for stable cancellation~\cite{Hua1}. Other analog cancellation of generating an RF signal with an auxiliary transmit chain was also introduced in~\cite{Duarte2014, Khandani1}, where the RF signal from the auxiliary chain was combined with the received signal to suppress self-interference. With a MIMO system, self-interference can be suppressed by transmit beamforming~\cite{Hua1}.

%%%%%%%%%%%%%%%%%%%%%%%%%%%%%%%%%%%%%%%%%%%%%%%%%%%%%%%%%%%%%%%%%%%%%%%%%%

\begin{figure*}[t!]
   \centerline{\resizebox{1.67\columnwidth}{!}{\includegraphics{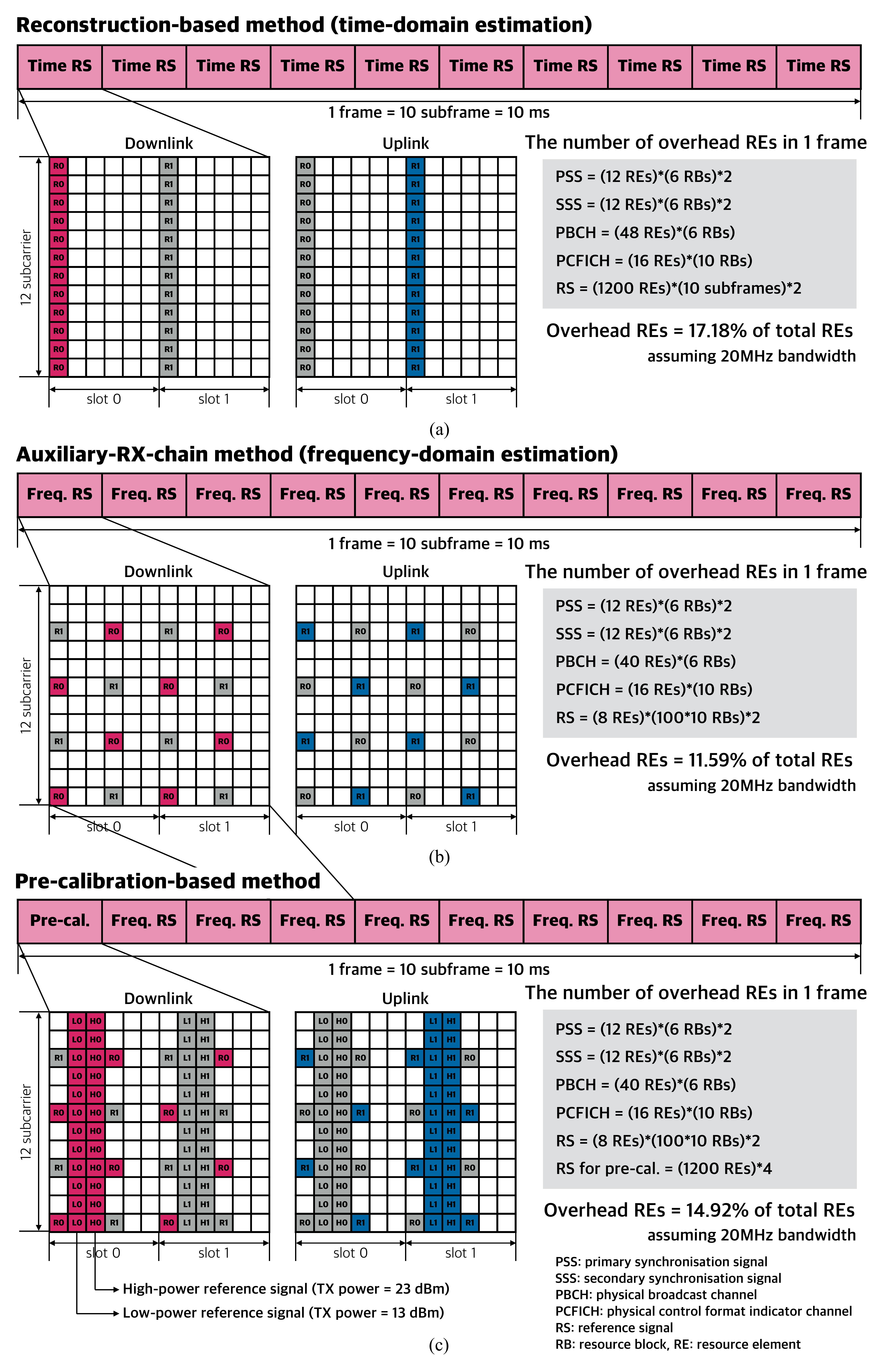}}}
   \caption{The frame structure, reference signal allocation, and the overhead in terms of resource-element ratio based on the LTE standard for each nonlinear digital self-interference cancellation, (a) the reconstruction-based method, (b) the auxiliary-receive-chain method, and (c) the pre-calibration-based method.}
   \label{fig.RS}
\end{figure*}

\begin{figure*}[t!]
   \centerline{\resizebox{1.5\columnwidth}{!}{\includegraphics{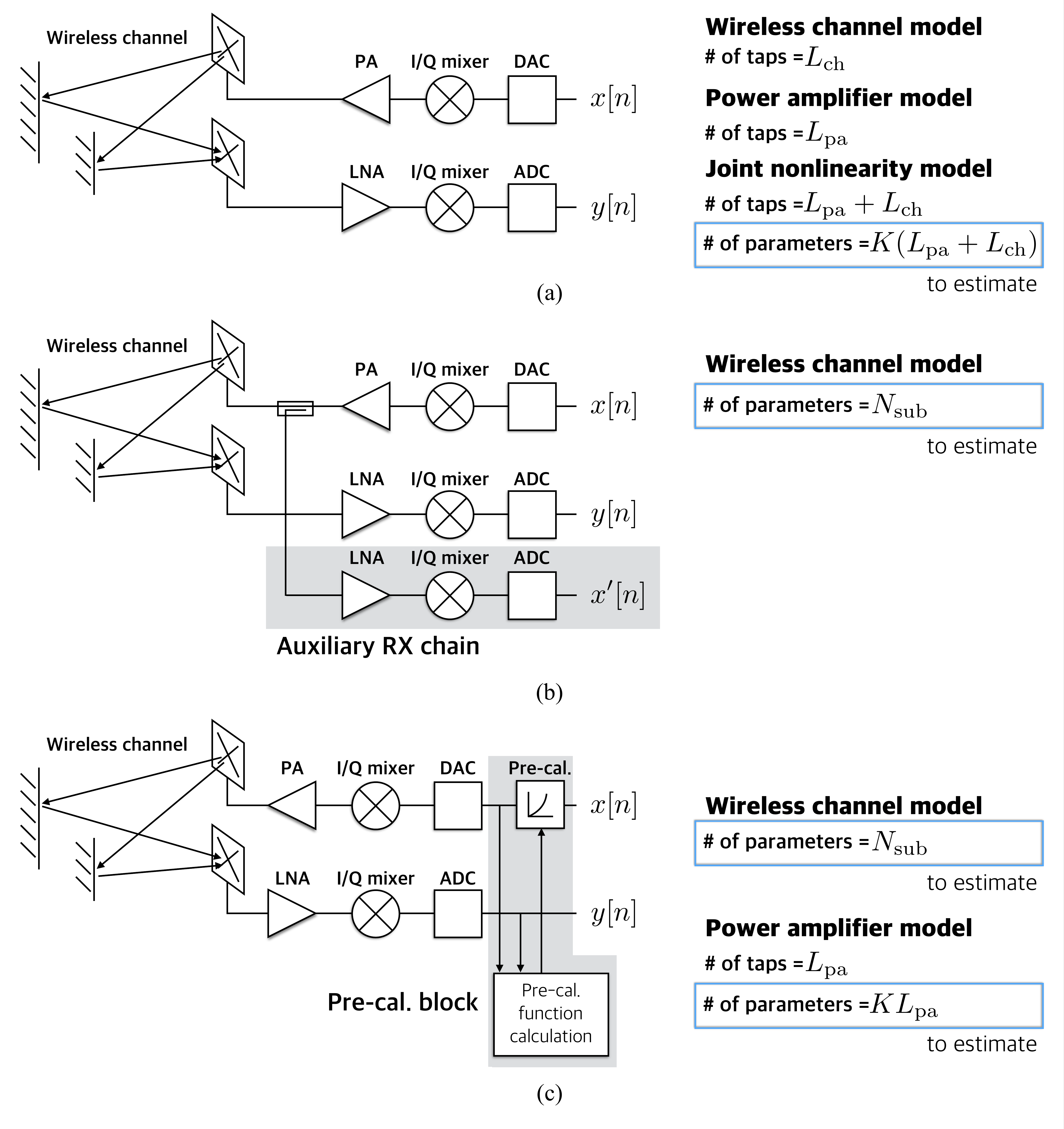}}}
   \caption{The schematic block diagram of each nonlinear digital self-interference cancellation method, and the number of parameters constructing each model, (a) the reconstruction-based method, (b) the auxiliary-receive-chain method, and (c) the pre-calibration-based method.}
   \label{fig.NDSIC}
\end{figure*}

\subsection{Linear Digital Self-Interference Cancellation}
\label{digitalSIC}

Digital self-interference cancellation cancels out the residual self-interference from analog self-interference cancellation. Generally, a line-of-sight component or a direct leakage of a circulator is suppressed by analog self-interference cancellation, but non-line-of-sight components (reflections) are not. Therefore, these components should be removed by digital cancellation. There are three steps for digital cancellation: 1) set a model of a self-interference signal, 2) estimate the channel, and 3) reconstruct the self-interference signal and subtract it from a received signal. Note that, in digital self-interference cancellation, a reference signal is needed not only for a self-interference, but also for a signal-of-interest.
% Figure~\ref{fig.RS} introduces three examples of reference signal allocation, and their overhead, calculated in terms of resource elements based on the LTE standard. 

Many researchers use, due to its simplicity, a linear model for a wireless channel. Therefore, self-interference channels are assumed to be linear, and linear self-interference cancellation becomes the basic technology of digital self-interference cancellation. In this article, we introduce two kinds of linear estimation methods and the following reference signal allocations. 
For simplicity, we assume orthogonal frequency-division multiplexing (OFDM) with the extended cyclic prefix (CP).

\smallskip
\subsubsection{Linear Digital Cancellation in Time Domain}

The first method is to estimate the self-interference channel in the time domain~\cite{Sachin2013Full}. Consider a full-duplex system that adopts, as shown in Fig.~\ref{fig.RS}(a), one OFDM symbol as a reference signal. In this case, the self-interference channel is obtained by the least squares method in the time domain. Note that this method should exploit the reference signal allocated in all subcarriers. To overcome the Doppler effect, however, the reference signals should be repeated, which creates a tremendous overhead for the reference signal. Furthermore, since the channel-estimation part includes pseudo-inverse operations, and the self-interference reconstruction part includes convolution operations, this method has high computational complexity.

\smallskip
\subsubsection{Linear Digital Cancellation in Frequency Domain}

It is more natural to allocate the reference signal as shown in Fig.~\ref{fig.RS}(b)~\cite{Chung2015Prototyping}, which is a structure similar to that used in the LTE standard. This reference signal is used for both the self-interference link and the desired link. With this pattern, the self-interference channel can be calculated by the least squares method in the frequency domain. In other words, the channel is obtained by dividing a received signal passed through a fast Fourier transform (FFT) block by the reference signal before an inverse FFT (IFFT) block. This method has relatively low complexity since it only requires element-wise multiplications and divisions, and interpolations.

%%%%%%%%%%%%%%%%%%%%%%%%%%%%%%%%%%%%%%%%%%%%%%%%%%%%%%%%%%%%%%%%%%%%%%%%%%

\section{Nonlinear Self-Interference Cancellation}
\label{nonlinear_SIC}

The need to cancel nonlinearity arises when we attempt to apply full-duplex radios to systems with high transmission power. Suppose that current LTE systems support a transmission power of 23~dBm. With analog cancellation that suppresses self-interference by approximately 60~dB~\cite{Sachin2013Full, Chung2015Prototyping, Heino2015FDRelay}, digital cancellation has the burden of self-interference to cancel of 50~dB ($\approx$~23~dBm~-~(-90~dBm)~-~60~dB). Digital cancellation, however, is limited by intermodulation distortion (IMD), caused by a power amplifier. Therefore, several techniques, called nonlinear digital self-interference cancellation, have been proposed to cancel the self-interference with IMD. 

\subsection{Nonlinearity Models}

As explained in Section~\ref{digitalSIC}, the first step in digital self-interference cancellation is determining what model to adopt. For decades, a challenging problem has been overcoming the nonlinearity of power amplifiers. Several theoretical approaches for modeling the nonlinearity have been developed such as the Wiener model and the Hammerstein model. In this article, we introduce the parallel Hammerstein model adopted in prior work~\cite{Sachin2013Full, Heino2015FDRelay}. The parallel Hammerstein model is expressed as:
$$y[n] = \sum_{k=0}^{K-1} \sum_{\ell=0}^{L-1} b_{k,\ell} |x[n-\ell]|^{2k} x[n-\ell] \text{,}$$
where $x[n]$ and $y[n]$ are the power amplifier's time-domain complex baseband input and output signals, $\{b_{k,\ell}\}$ are the coefficients of the model, and $2K-1$ and $L$ represent the order of the model and the number of the model's taps. Note that the number of $\{b_{k,\ell}\}$ is $KL$. This model is constructed of odd-order terms of the input signal because in wireless communication systems the only thing considered is a passband signal near a center frequency. It can be inferred that the model becomes memoryless if $L=1$, and linear if $K=1$.

Expressed in terms of the complex baseband signals, the parallel Hammerstein model can be affected by I/Q imbalances. The effect of I/Q imbalances can be avoided by employing the real-valued model introduced in~\cite{Hua1}. The authors in~\cite{Heino2015FDRelay} introduced the polynomial basis function, which includes the effect of I/Q imbalances. In this article, we assume that the system follows the parallel Hammerstein model, which can be directly applied to LTE systems.

\subsection{Conventional Nonlinear Cancellation Techniques}
\label{subsec.nonlinear_SIC}

In contrast to canceling linear components, canceling nonlinear components calls for extra resources such as hardware, pilot overhead, and/or computational complexity. In this section, we introduce two representative conventional nonlinear digital-cancellation techniques, and introduce their strengths and weaknesses. Figure~\ref{fig.NDSIC} shows the schematic block diagrams and the number of parameters constructing each model. $x'[n]$ is the measured signal via an auxiliary receive chain. $L_\text{ch}$ is the delay spread of a self-interference channel, $L_\text{pa}$ is the number of memory taps of a power amplifier, and $N_\text{sub}$ is the number of used subcarriers.

\smallskip
\subsubsection{Reconstruction-based Method}

This method follows a typical digital cancellation methodology; estimate and reconstruct the nonlinear self-interference signal, and subtract it from the received signal~\cite{Sachin2013Full, Heino2015FDRelay}. To estimate the nonlinearity---similar to linear digital cancellation in the time domain---the reference signal allocation shown in Fig.~\ref{fig.RS}(a) and the least squares method are adopted. Two signals used to estimate coefficients are the signal before passing a power amplifier, and the received signal. Unlike linear channel estimation, the model of self-interference signal is reformulated into odd-order terms, and the number of the coefficients to be estimated is increased $K$-fold. The larger number of coefficients causes more complex computation to estimate them. Furthermore, if the number of coefficients is greater than the number of the samples of the reference signal, it makes the least squares problem something that should be avoided---an undetermined system.
% To solve this problem, we need to neglect some of the high-order coefficients, such that the power of $b_{k,\ell} |x[n-\ell]|^{2k} x[n-\ell]$ is much smaller than the noise level~\cite{Sachin2013Full}. These coefficients are chosen before self-interference cancellation by measurement.

The major weakness of this method is that the nonlinearity model consists of both a power amplifier and a wireless self-interference channel. This is why the number of the coefficients is abnormally large. Figure~\ref{fig.NDSIC}(a) shows the number of parameters of the power amplifier, wireless channel, and the joint nonlinearity model. The taps of the nonlinearity model are caused not only by the memory effect of the power amplifier, but mostly by the reflections of the wireless channel. Therefore, the number of coefficients is large, and real-time cancellation becomes difficult due to computational complexity. Furthermore, the wireless channel makes the total nonlinear model time-varying. In other words, the nonlinear model needs to be estimated more frequently, thus the overhead of the reference signal increases. One might argue that the overhead can be reduced by utilizing the data signal, which is always known to the transceiver, as a reference signal. However, since the reference signal is also needed for another node which is trying to receive the signal, we can consider as if there is no extra reference signal for self-interference cancellation.

\smallskip
\subsubsection{Auxiliary-Receive-Chain Method}

The nonlinear self-interference can be canceled by using an auxiliary receive chain~\cite{Ahmed2015All}. Through the auxiliary receive chain, as illustrated in Fig.~\ref{fig.NDSIC}(b), the signal distorted by a power amplifier is directly obtained rather than estimated or reconstructed. This distorted signal is used to estimate the self-interference channel and to reconstruct the self-interference signal. Recall that conventional linear digital self-interference cancellation methods obtain the reference signal before a power amplifier distortion. The limitation of such a method is that the self-interference cannot be reconstructed accurately since it is assumed to pass through a linear system, which is in fact nonlinear. On the other hand, by using the power-amplifier-distorted signal as the reference signal, it can be assumed that this signal passes through a linear system---the wireless channel. The estimation and reconstruction schemes are the same as with linear self-interference cancellation, and both the time-domain and frequency-domain processing can be applied. Therefore, there is no need for extra computational resources compared to linear cancellation. The only drawback of this method is the need for the extra receive chain.

\subsection{Proposed Pre-calibration-based Cancellation Method}
\label{sec.precal_SIC}

To cancel nonlinear self-interference, we propose a pre-calibration-based cancellation technique which linearizes a transmitter and cancels self-interference with linear-only cancellation at a receiver. A pre-calibrator, as illustrated in Fig.~\ref{fig.NDSIC}(c), estimates the nonlinearity of a power amplifier, and modifies the input signal of the power amplifier to linearize the output signal of the power amplifier. Put simply, since the signal with a high amplitude is saturated by the power amplifier, a pre-calibrator strengthens the high-amplitude-signal more than the low-amplitude-signal. Then, the combined system of the pre-calibrator and the power amplifier is linearized. As IMD---which limits the linear self-interference cancellation---decreases, the linearized self-interference signal can be canceled by the linear digital self-interference cancellation technologies. Furthermore, a receiver can achieve higher signal-to-noise ratio (SNR) with improved transmit error vector magnitude (EVM).

\smallskip
\subsubsection{Frame Structure of Reference Signal}
A frame structure of a reference signal, as illustrated in Fig.~\ref{fig.RS}(c), consists of two parts. The first part is for calculating a pre-calibrator. This reference signal is for measuring the characteristic of the power amplifier. Since the characteristic of the power amplifier is more static than that of a wireless channel, this reference signal can be allocated with a long period compared to the coherence time of the wireless channel. For instance, in this article, we allocate this reference signal every 10~ms (1 frame in the LTE standard) as illustrated in Fig.~\ref{fig.RS}(c). A detail of this reference signal is explained in a later section. The second part is for data transmission and linear cancellation. Due to its low complexity and low overhead explained in Section~\ref{digitalSIC}, the reference signal allocation shown in Fig.~\ref{fig.RS}(b), and linear cancellation in the frequency domain are adopted.

\smallskip
\subsubsection{Calculating a Pre-calibrator}
\label{subsec.precal_cal}

To measure the output of the power amplifier without an extra wire, we propose the following technique. To model the power amplifier, two signals are needed---the input signal, and the output signal of the power amplifier. Note that what we need is the output signal of the power amplifier that does not pass through wireless channel. Therefore, we decide to exploit the accurately estimated wireless channel and remove its effect from the received signal by equalization. We suggest the reference signal containing a low-power signal and a high-power signal, as illustrated in Fig.~\ref{fig.RS}(c). The low-power reference signal (denoted by `L0') aims for accurate wireless channel estimation. So, to avoid distortion, it operates on a linear (low-power) region of the power amplifier, and to lessen estimation errors, it occupies all subcarriers. The high-power reference signal (denoted by `H0')  is for estimating the power amplifier's distortion. This signal experiences a nonlinear (high-power) region of the power amplifier, and undergoes sufficient distortion. Then, using the precisely estimated wireless channel, the output of the power amplifier can be calculated.

There are several conventional methods to calculate a pre-calibration function. The method used in this article is to estimate coefficients of a polynomial-based pre-calibration function. It is just like estimating a nonlinearity model, but uses the input of the power amplifier as output and vice versa to estimate a reversed function. The pre-calibration function is obtained from the reversed function of the power amplifier, and scaling the power amplifier gain.

\smallskip
\subsubsection{Strengths of Proposed Method}
The proposed method has two main strengths compared to the conventional methods. 
First, there is no need for extra receive chain or wire. Techniques for power amplifier linearization have been applied to most conventional radios with high power. One of the effective technologies to reduce the nonlinearity of a transmitter is digital pre-distortion (DPD)~\cite{Kim2005DPD}. However, most DPD systems need a secondary receive chain to estimate the output of the power amplifier~\cite{Kim2005DPD}. In~\cite{Khandani1}, a full-duplex system which measures the effect of the power amplifier through a cable, and applies to an auxiliary transmit signal was proposed. With full-duplex systems, since a receiver is able to sense a transmit signal without saturation, a pre-calibration function can be calculated without extra receive chain and cable.

Second, complex and time-varying nonlinear cancellation is not required. Contrary to the reconstruction-based method, the pre-calibration-based method takes advantages of the nonlinear model of the power amplifier without the effect of wireless channels. It is, therefore, expected that the coefficients $\{b_{k,\ell}\}$ are more static and fewer in number, which lessens the burden of computational complexity and the reference-signal overhead. For example, with the proposed reference signal illustrated in Fig.~\ref{fig.RS}(c), as shown in Fig.~\ref{fig.NDSIC}, $KL_\text{pa}$ coefficients need to be estimated every 10~ms, while the reconstruction-based method estimates $K(L_\text{pa}+L_\text{ch})$ coefficients every 1~ms.

\begin{figure*}[t!]
   \centerline{\resizebox{2.0\columnwidth}{!}{\includegraphics{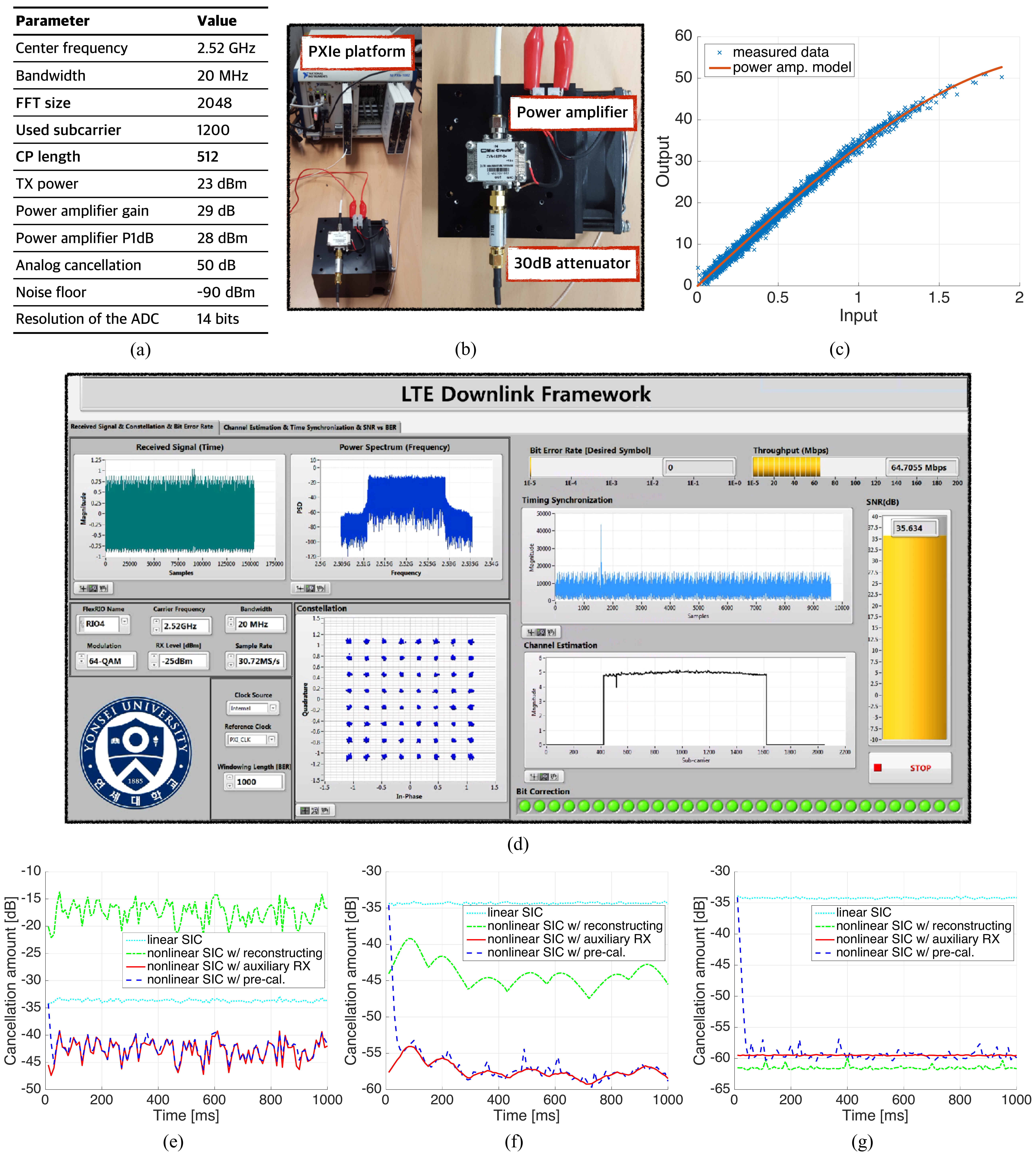}}}
   \caption{(a) The parameters for link- and system-level simulations, (b) on the left side is a setup for measuring the characteristic of the power amplifier, and on the right side is the used power amplifier and the attenuator, (c) measured input and output signals of the power amplifier, and an approximated model of the power amplifier, (d) the GUI of the LTE downlink framework used for measuring the model of the power amplifier, (e), (f) the results of the link-level simulations with the coherence time of 7.13~ms, and 142.86~ms, and (g) the result of the link-level simulation with constant wireless channel.}
   \label{fig.link_level}
\end{figure*}

%%%%%%%%%%%%%%%%%%%%%%%%%%%%%%%%%%%%%%%%%%%%%%%%%%%%%%%%%%%%%%%%%%%%%%%%%%

\section{Performance Evaluations}
\label{evaluation}

We evaluate the nonlinear digital self-interference cancellation techniques from two points of view: link-level evaluations for measuring nonlinearity of a power amplifier and cancellation amounts, and system-level evaluations for throughput gain. A basic criterion of self-interference cancellation technologies is the amount of cancellation. This is important because it guarantees the feasibility of full-duplex radios. 

Simple link-level evaluations, however, do not fully explain why nonlinear digital cancellation is needed despite its high complexity. Most prior work has tried to suppress self-interference to the same level as the noise floor, but it is questionable whether this is really necessary in practice where there is other interferences. Therefore, we also carry out, with a 3D ray-tracing tool, system-level simulations of multi-BSs and multi-mobile stations (MSs) in an indoor environment. The link-/system-level analyses were based on computer simulations. To obtain a realistic power amplifier model, we measured the characteristics of a power amplifier via an LTE-based software-defined radio platform that we developed. To model a 3D indoor environment and obtain path-losses between nodes, a 3D ray-tracing tool was employed.

\begin{figure*}[t!]
   \centerline{\resizebox{2.0\columnwidth}{!}{\includegraphics{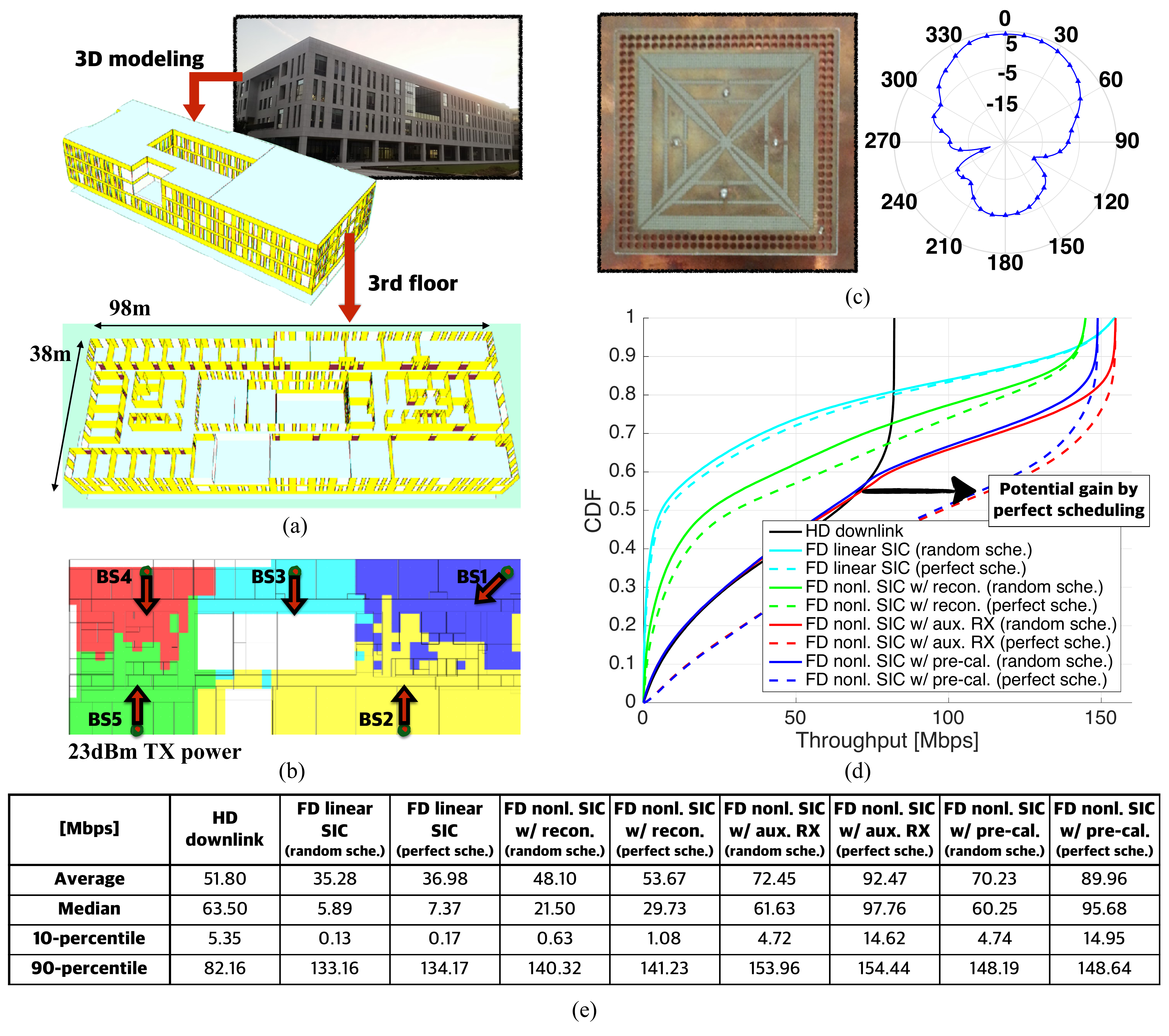}}}
   \caption{(a) Topology for system-level performance evaluations, (b) BS deployments and cell coverages, (c) dual-polarized antenna and radiation pattern, (d) CDF of the system throughput, and (e) throughput results. Here, `random scheduling' means random user selection and `perfect scheduling' means Genie-aided ideal user selection.}
   \label{fig.system_level}
\end{figure*}

\subsection{Link-Level Self-Interference Cancellation Performance}
\label{link_simul}

To simulate nonlinear digital cancellation, we assume a single-input and single-output (SISO) wireless system based on the LTE parameters given in Fig.~\ref{fig.link_level}(a). 
To exploit a realistic model for a power amplifier, we measure the nonlinearity of a power amplifier\footnote{Mini-Circuits ZVA-183W+ Super Ultra Wideband Amplifier, http://www.minicircuits.com/pdfs/ZVA-183W+.pdf} through the PXIe software-defined radio platform\footnote{With this platform, we developed and demonstrated the real-time full-duplex SISO and MIMO systems in IEEE GLOBECOM~2014 and IEEE GLOBECOM~2015, respectively. Full demo video clips are available at http://www.cbchae.org/} introduced in~\cite{Chung2015Prototyping}. This platform generates an LTE-based signal with given physical-layer parameters. As shown in Fig.~\ref{fig.link_level}(b), the input and output (distorted) signals of the power amplifier are measured via the PXIe platform. A 30~dB attenuator is employed to receive the output signal of the power amplifier without distortion from the receiver of the PXIe platform. The GUI of the LTE downlink framework, which is shown in Fig.~\ref{fig.link_level}(d), shows the received signal has passed through the power amplifier. Figure~\ref{fig.link_level}(c) shows the measured input and output signals of the power amplifier, and a parallel-Hammerstein-model-based approximated power amplifier model for simulations. In this simulation, the power amplifier was modeled as the third order model, which showed the greatest similarity with the measured data. Tending to make the residual self-interference channel frequency selective, analog self-interference cancellation is simulated by employing a longer-than-normal delay-spread for the residual self-interference channel and attenuating the channel 50~dB.

The three nonlinear digital cancellation techniques introduced in Section~\ref{subsec.nonlinear_SIC} and Section~\ref{sec.precal_SIC} were evaluated, and compared with a frequency-domain linear digital cancellation. The reference signal of the reconstruction-based method was determined to be repeated every subframe (12~OFDM symbols) considering the Doppler effect and the reference signal overhead, as illustrated in Fig.~\ref{fig.RS}(a). For the linear cancellation part of the auxiliary-receive-chain method and the pre-calibration-based method, the frequency-domain-based linear cancellation was adopted, and the reference signal pattern followed the cell-specific reference signal in LTE, as seen in Fig.~\ref{fig.RS}(b). The orders of the reconstructed parallel Hammerstein model and the pre-calibration function are 7 and 5, respectively.

We simulated the amount of cancellation with different coherence times. Figures~\ref{fig.link_level}(e),~and \ref{fig.link_level}(f) show how much self-interference power was reduced when the coherence times of the self-interference channel were 7.14~ms, and 142.86~ms, respectively. These coherence times were calculated from the Doppler effect, assuming that the center frequency was 2.52~GHz, and MSs were vehicles (60~km/h) or pedestrians (3~km/h). Figure~\ref{fig.link_level}(g) shows the amount of self-interference cancellation with a static self-interference channel.

When the self-interference channel was static, the performance of the reconstruction-based method, as Fig.~\ref{fig.link_level}(g) shows, was about 62~dB---the best performance among the three nonlinear cancellation methods. This occurred because this method can reconstruct self-interference accurately (only limited by noise), but the other two methods were limited by channel estimation error from the interpolating process in the frequency domain. On the other hand, it is observed that its cancellation performance is 43~dB when the coherence time was 142.86~ms, and when the coherence time was 7.13~ms, even worse than that of linear self-interference cancellation. Therefore, we conclude that in practice it is not proper to cancel self-interference by reconstructing, due to its weakness in the time-varying channel.

The auxiliary-receive-chain method and the pre-calibration-based method are robust to fading channel. Figure~\ref{fig.link_level}(e) shows that, even though the coherence time was small, these methods could cancel nonlinear self-interference by about 43~dB. This result is quite obvious because the reference signals exploit the very same structure of LTE. The performances of these two methods degrade as coherence time decreases because of an increase in the channel estimation error from the interpolating process in the time domain. The auxiliary-receive-chain method behaves as a performance-upper-bound of the pre-calibration-based method because the pre-calibrator does not linearize ideally. One might argue that the performance of the pre-calibration-based method is not as stable as that of the auxiliary-receive-chain method. This is because the pre-calibration function generator, which is calculated in every frame, fails to generate the well-operating pre-calibration function in some frames. By applying adaptive polynomial filters~\cite{John1991adaptive}, or orthogonal polynomials~\cite{Raich2004orthogonal}, we can have better chance of improving performance. We leave this for future work.

\subsection{System-Level Throughput Gain}

We investigated the need for nonlinear digital self-interference cancellation technologies from a system throughput perspective. For system-level evaluations, we used Wireless System Engineering (WiSE)---a 3D ray-tracing tool developed by Bell Labs~\cite{Valenzuela1998wise, Jang2015Smart}. For an indoor environment, we modeled the building structure of Veritas Hall C of Yonsei University in Korea, shown in Fig.~\ref{fig.system_level}(a). Each BS is modeled with the measured radiation pattern of the dual-polarized antenna introduced in~\cite{Oh2014, Chung2015Prototyping} (see Fig.~\ref{fig.system_level}(c)). On the third floor, there were uniformly distributed MSs, each of which was equipped with an isotropic antenna; these were associated with the BS that provided the strongest downlink power. Figure~\ref{fig.system_level}(b) illustrates how five BSs were deployed on the third floor, and how each cell coverage was determined. Note that due to its radiation pattern, each BS has its own direction, represented as an arrow in Fig.~\ref{fig.system_level}(b). For evaluations, we assumed that there were five MSs, which were located in different cell coverages, and adopted the parameters in Fig.~\ref{fig.link_level}(a).

In system-level simulations, we investigated the throughput of a half-duplex downlink system based on FDD LTE, and full-duplex systems with different self-interference cancellation levels. For the cancellation performances, we exploited the average of the link-level simulation results with the coherence time of 142.86~ms in Section~\ref{link_simul}, and combined them with analog cancellation of 50~dB given in Fig.~\ref{fig.link_level}(a). The exact cancellation amounts were 84.28~dB for linear cancellation, 93.03~dB for reconstruction-based nonlinear cancellation, 107.17~dB for auxiliary-receive chain nonlinear cancellation, and 106.20~dB for pre-calibration-based nonlinear cancellation. With the received power of signal-of-interest and the summation of the received power of interferences and noise, signal-to-interference-plus-noise-ratio (SINR) of each node was calculated. Finally, a system bandwidth of 20~MHz and the overheads introduced in Fig.~\ref{fig.RS} were applied. Note that there was also the overhead for the extended CP. We also simulated the full-duplex systems without considering MS-to-MS interference as upper bounds, which could be achieved by Genie-aided perfect scheduling.

Figure~\ref{fig.system_level}(d) illustrates the results of the ergodic throughput of the half-duplex system and the full-duplex systems with different self-interference levels. The solid lines of full-duplex indicate results with MS-to-MS interference and the dashed lines indicate those without it. Figure~\ref{fig.system_level}(e) gives the representative values of each case such as average, median, 10-percentile, and 90-percentile. The result implies that the throughput of a full-duplex system depends heavily on the performance of self-interference cancellation. If the full-duplex system employs only linear self-interference cancellation, the average throughput is lower than that of the half-duplex system, and only about 20\% of MSs can experience the benefit of full duplex. On the other hand, with the well-performing nonlinear self-interference cancellation techniques such as the auxiliary-receive-chain method or the pre-calibration-based method, the average throughput increased by approximately 35 to 40\% over that of a half-duplex system. Even though self-interference is canceled out sufficiently, we can observe that for certain portions of MSs, half-duplex outperforms full-duplex. Furthermore, if the interference between MSs was avoided somehow, the average throughput could be improved by up to 79\%. From this, we have the insight that once self-interference cancellation techniques guarantee certain performances, a critical issue in full-duplex research will become user-allocation and user-scheduling.

\section{Conclusion}
\label{conclusion}

%Full-duplex radio is expected to play a major role to further maximize the spectral efficiency in 5G wireless communications/LTE Evolution. In this article, we first explored several analog self-interference cancellation and linear digital cancellation methods. We then investigated nonlinear digital cancellation methods to understand the importance of which keeps increasing where systems exploit wide bandwidth and high power. We evaluated the nonlinear digital cancellation schemes from two points of view: the amount of cancellation via the link-level simulation, and system data rate via the system-level simulation. From the link-level simulation, we investigated how reference-signal allocation influenced the cancellation performance in a practical fading channel. The system-level evaluation highlighted the importance of high-performance self-interference cancellation and user scheduling in an indoor environment. We expect our study to provide insights into developing practical cellular systems based on full-duplex radio.

Full-duplex radio is expected to play a major role in enhancing the spectral efficiency in 5G wireless communications/LTE Evolution. In this article, we have investigated two existing nonlinear digital cancellation techniques and proposed a low complexity pre-calibration-based technique. Link-level and system-level performances were analyzed through a real-time software-defined radio platform and a 3D-ray-tracing-based simulations of an indoor environment. The results of our analysis confirmed a significant performance enhancement even in interference-limited environments. We expect our study to provide insights into developing practical cellular systems based on full-duplex radios.

% use section* for acknowledgement
\section*{Acknowledgment}
The authors would like to thank Mr. Y.-G. Lim for helpful discussions on WiSE simulations.

% Can use something like this to put references on a page
% by themselves when using endfloat and the captionsoff option.

\ifCLASSOPTIONcaptionsoff
  \newpage
\fi

\renewcommand{\baselinestretch}{1.0}
\bibliographystyle{IEEEtran}
\bibliography{reference_CommMag15} % file name

\begin{IEEEbiography}{Min Soo Sim}(S'14)
received his B.S. degree in the School of Integrated Technology from Yonsei University, Korea, in 2014. He is now with the School of Integrated Technology, at the same university and is working toward a Ph.D. degree. He was the recipient of the Silver Prize in the 22nd Humantech Paper Contest. His research interest includes emerging technologies for 5G communications. 
\end{IEEEbiography}
\begin{IEEEbiography}{MinKeun Chung}(S'11)
received his B.S. degree in the School of Electrical and Electronic Engineering from Yonsei University, Korea, in 2010. He is now working toward the Ph.D. degree under the joint supervision of Prof. D. K. Kim and Prof. C.-B. Chae. He did his graduate intern in advanced wireless research team at the National Instruments in Austin, TX, USA in 2013. His research interests include the design and implementation of architectures for next-generation wireless communication systems.
\end{IEEEbiography}
\begin{IEEEbiography}{Dongkyu Kim}(M'13)
received his B.S. degree in electrical engineering from Konkuk University, Seoul, South Korea, in 2006 and his M.S. and Ph.D. degrees in electrical and electronic engineering from Yonsei University, Seoul, in 2008 and 2013, respectively. From 2013 to 2014, he was a Post-Doctoral Researcher with the Information and Telecommunication Laboratory, Yonsei University. Since 2014, he has been with LG Electronics Inc. as a Senior Researcher, where he was involved in development of advanced wireless technologies, including 5G mobile communications and 3GPP standard for future wireless systems. His current research activities are focused on future wireless communication including flexible and full-duplex radio, V2X, massive MIMO, and mmWave technologies.
\end{IEEEbiography}
\begin{IEEEbiography}{Jaehoon Chung}
is with LG Electronics Inc., where he was involved in development of advanced wireless technologies, including 5G mobile communications and 3GPP standard for future wireless systems. %His current research activities are focused on future wireless communication including flexible and full-duplex radio, V2X, massive MIMO, and mmWave technologies.
\end{IEEEbiography}
\begin{IEEEbiography}{Dong Ku Kim}(SM'15)
received his B.S. from Korea Aerospace University in 1983, and his M.S. and Ph.D. from the University of Southern California, Los Angeles, in 1985 and 1992, respectively. He worked on CDMA systems in the cellular infrastructure group of Motorola at Fort Worth, Texas, in 1992. He has been a professor in the School of Electrical and Electronic Engineering, Yonsei University since 1994. Currently, he is a vice president for academic research affairs of the Korean Institute of Communications and Information Systems (KICS). He has been a vice chair of the executive committee of 5G Forum since 2013. He received the Minister Award for the Distinguished Service for ICT R\&D from the Ministry of Information, Science, and Future Planning in 2013, and the Award of Excellence in leadership of 100 Leading Core Technologies for Korea 2020 from the National Academy of Engineering of Korea. He received  Dr. Irwin Jacobs Academic Achievement Award 2016 from Qualcomm and KICS.
\end{IEEEbiography}
\begin{IEEEbiography}{Chan-Byoung Chae}(SM'12)
is an associate professor in the School of Integrated Technology, Yonsei University. Before joining Yonsei University, he was with Bell Labs, Alcatel-Lucent, Murray Hill, New Jersey, as a member of technical staff, and Harvard University, Cambridge, Massachusetts, as a postdoctoral research fellow. He received his Ph.D. degree in electrical and computer engineering from the University of Texas at Austin in 2008. He was the recipient/co-recipient of the IEEE INFOCOM Best Demo Award (2015), the IEIE/IEEE Joint Award for Young IT Engineer of the Year (2014), the KICS Haedong Young Scholar Award (2013), the IEEE Signal Processing Magazine Best Paper Award (2013), the IEEE ComSoc AP Outstanding Young Researcher Award (2012), the IEEE Dan. E. Noble Fellowship Award (2008), and two Gold Prizes (1st) in the 14th/19th Humantech Paper Contest. He currently serves as an Editor for IEEE Transactions on Wireless Communications, the IEEE/KICS Journal on Communications Networks, and IEEE Transactions on Molecular, Biological, and Multi-scale Communications.
\end{IEEEbiography}

% You can push biographies down or up by placing
% a \vfill before or after them. The appropriate
% use of \vfill depends on what kind of text is
% on the last page and whether or not the columns
% are being equalized.

%\vfill

% Can be used to pull up biographies so that the bottom of the last one
% is flush with the other column.
%\enlargethispage{-5in}

% that's all folks
\end{document}